\documentclass[number,preprint,review,12pt]{elsarticle}
\biboptions{sort&compress}
\usepackage{lineno}
\usepackage[pdftex,plainpages=false,pdfpagelabels,colorlinks=true,citecolor=Blue,linkcolor=DarkBlue,urlcolor=DarkRed,filecolor=green,bookmarksopen=true]{hyperref}

\def\txtcolorSec{black}

\usepackage[utf8]{inputenc}
\usepackage[top=1in,bottom=1in,left=1in,right=1in]{geometry}
\usepackage{amsmath,amssymb,amsfonts,amsthm,accents,mathrsfs,bm}
\usepackage{graphicx,subcaption}
\usepackage{wrapfig,epsfig}
\usepackage{setspace}         
\usepackage{comment}
\usepackage{enumerate,makeidx}
\usepackage{dcolumn}          
\usepackage{booktabs,colortbl,array}
\usepackage{sectsty,lastpage}
\usepackage[displaymath,textmath,sections,graphics,floats]{preview}
\PreviewEnvironment{enumerate,tabular}
\usepackage{hyphenat}
\usepackage{nicematrix}
\usepackage{soul}
\usepackage{algorithm, algpseudocode}
\DeclareCaptionLabelFormat{myformat}{\textcolor{\txtcolorSec}{Code block #2}}
\usepackage{multicol,multirow} 
\usepackage{xstring}
\newcommand{\textuser}[2]{%
    \IfEqCase{#1}{
        {1}{\textit{#2}}
        {2}{\texttt{#2}}
        {3}{\textnormal{#2}}
    }[\PackageError{textuser}{Undefined option to textuser: #1}{}]%
}%
\usepackage{csquotes}

\journal{International Journal of Mechanical Sciences}

\biboptions{sort&compress}

\begin{document}
\begin{frontmatter}
  \title{Graphene Design with Parallel Cracks: Abnormal Crack
    Coalescence and Its Impact on Mechanical Properties}

\author[mymainaddress1,mymainaddress2]{Suyeong Jin}
\author[mymainaddress1]{Jung-Wuk Hong}
\author[mymainaddress2]{Chiara Daraio\corref{mycorrespondingauthor2}}
\cortext[mycorrespondingauthor2]{Corresponding author}\ead{daraio@caltech.edu}
\author[mymainaddress2,mymainaddress3]{Alexandre F. Fonseca\corref{mycorrespondingauthor1}}
\cortext[mycorrespondingauthor1]{Corresponding author}\ead{afonseca@ifi.unicamp.br}

\address[mymainaddress1]{Department of Civil and Environmental Engineering, Korea Advanced Institute of Science and Technology, 291 Daehak-ro, Yuseong-gu, Daejeon 34141, Republic of Korea}
\address[mymainaddress2]{Division of Engineering and Applied Science, California Institute of Technology, Pasadena, CA 91125, USA}
\address[mymainaddress3]{Universidade Estadual de Campinas, Instituto de F\'{i}sica Gleb Wataghin, Departamento de F\'{i}sica Aplicada, 13083-859, Campinas, SP, Brazil}

\begin{abstract}
Graphene is a material with potential applications in electric,
thermal, and mechanical fields, and has seen significant advancements
in growth methods that facilitate large-scale production. However,
defects during growth and transfer to other substrates can compromise
the integrity and strength of graphene. Surprisingly, the literature
suggests that, in certain cases, defects can enhance or, at most, not
affect the mechanical performance of graphene. Further research is
necessary to explore how defects interact within graphene structure
and affect its properties, especially in large-area samples.  In this
study, we investigate the interaction between two preexisting cracks
and their effect on the mechanical properties of graphene using
molecular dynamics simulations. The behavior of zigzag and armchair
graphene structures with cracks separated by distances
($W_\text{gap}$) is analyzed under tensile loading. The findings
reveal that crack coalescence, defined as the formation of a new crack
from two existing crack tips, occurs for lower values of the distance
between cracks, $W_\text{gap}$, resulting in a decline in the strength
of structures.  As $W_\text{gap}$ increases, the stress-strain curves
shift upward, with the peak stress rising in the absence of crack
coalescence. The effective stress intensity factor formulated in this
study exhibits a clear upward trend with increasing $W_\text{gap}$.
Furthermore, an increase in $W\text{gap}$ induces a transition in
fracture behavior from crack coalescence to independent propagation
with intercrack undulation. This shift in fracture behavior
demonstrates a brittle-to-ductile transition, as evidenced by
increased energy absorption and delayed failure.  A design guideline
for the initial crack geometry is suggested by correlating peak stress
with the $W_\text{gap}$, within a certain range.  The findings offer
insights into the fracture mechanics of graphene, emphasizing the
impact of crack interaction and geometry on strength. This provides
design guidelines for graphene-based structures with enhanced
mechanical performance.
\end{abstract}
\begin{keyword}
Graphene design\sep Parallel cracks\sep Crack coalescence \sep Fracture \sep Stress intensity factor
\end{keyword}

\end{frontmatter}

\section{Introduction}
Graphene is a promising material for various electrical, thermal, and
mechanical applications because of its exceptional physical
properties~\citep{Novoselov2004Science,Neto2009RPG,Zhang2014,Papageorgiou2017}.
With a Young's modulus of 1~TPa and an intrinsic strength of
130~GPa~\citep{Changgu2008}, pristine graphene became the paradigm of
a lightweight material with remarkable mechanical properties.  Despite
these outstanding properties, the development of large scale
applications is still facing significant challenges in both the growth
of large-area perfect graphene and its transfer without damages to any
kind of substrate.

In fact, the production of large-area graphene has been possible for
more than a decade through chemical vapor deposition (CVD)
method~\citep{Xuesong2009}.  This technique exploits the low
solubility of carbon in metals such as copper~\citep{Levendorf2009}
and nickel~\citep{Yu2008} to facilitate graphene growth on metallic
foils.  Despite its popularity and widespread use, the CVD growth
method has not yet produced graphene samples of consistent quality and
reproducibility~\citep{Lin2009NatComm,Boggild2023NatComm}.  Although
one of these challenges is being surmounted (see, for example the
recent report of Amontree {\it et al.}~\citep{Amontree2024Nat}), it
still remains challenging to transfer as-grown graphene samples to
different substrates without damaging them~\citep{Boggild2023NatComm}.
The transfer process can introduce imperfections, such as cracks and
tears, into graphene films, which might affect their
properties~\citep{Pham2024}.

Notwithstanding the aforementioned challenges, it may be posited that
the imperfections of graphene may not be as deleterious as initially
presumed. For instance, graphene sheets with grain boundaries have
been demonstrated to exhibit strength that is comparable to that of
the pristine material~\citep{Rassin2010}. The ultimate failure of
nanocrystalline graphene with previous flaws was demonstrated to not
necessarily initiate at the flaw~\citep{Zhang2012}.  Defects have been
shown to potentially confine crack propagation in
graphene~\citep{Lopez-Polin2015}, and dislocations have been found to
be shielded on graphene nanocracks~\citep{Meng2015}.

While these and many other similar studies have contributed to the
understanding of the mechanical properties of graphene with and
without different kinds of defects and flaws, the study of
interactions {\it between} different flaws in graphene remains much
less explored.  A few examples can be given.  Dewapriya and
collaborators~\citep{DewaCarbon2017,DewaEngFracMech2018,DewaCMS2018}
computationally simulated the interaction between crack and vacancies
and/or holes, as well as boron-nitride inclusions in graphene.  Yao
{\it et al.}~\citep{YaoEngFracMech2019} have performed a similar study
of nanoscale crack-hole interactions in chiral graphene nanoribbons.
Inspired by an experimental work on asymmetric crack propagation in
bidimensional polymeric materials containing sequences of cracks with
a given geometry~\citep{BrodnikPRL2021}, Felix and
Galvao~\cite{felixPCCP2022} demonstrated through reactive molecular
dynamics simulations the same rectification effect in crack
propagation in graphene structures.  These works demonstrated that the
shape, size and position of the corresponding flaws have a significant
impact on the enhancement of the mechanical properties of graphene.

In this study, the coalescence of two cracks and its effects on the
fracture behavior of graphene are investigated as a function of crack
gap.  Computational tools of classical molecular dynamics (MD) with a
known reactive force field are employed to simulate the application of
tensile strains to the structures from the equilibrium to full
rupture.  Coalescence between cracks occurs during tensile loading in
graphene samples when the crack gap is below a certain value, leading
to deterioration in the material strength.  However, above that value,
the rupture of graphene displays ductile behavior, demonstrated by
energy absorption and the fracture pattern.  Since ductile behavior in
graphene has been ruled out at any temperature below the melting
point~\cite{Antonio2022CARBONTRENDS}, this last effect indicates that
the understanding of how two or more topological defects interact
within graphene can enable the design of brittle-to-ductile behavior
in this structure.

This work is organized as follows.  In section \ref{sec2:methodology},
we present the models of graphene samples with previous cracks, the
computational methods employed to investigate the crack coalescence in
graphene, and the results of three preliminary tests needed to
determine the best protocols for this research.  In
section~\ref{sec:main-ss-curve}, the main results and discussion of
the tensile strain numerical experiments are presented.  The main
conclusions of this study are presented in
section~\ref{sec:conclusion}.

\section{\label{sec2:methodology}Computational Models and Methods}

\subsection{\label{sec:graphene-geometry}Graphene structures and the geometry of preexisting cracks}
\begin{figure}
    \centering
    \includegraphics[width=\linewidth]{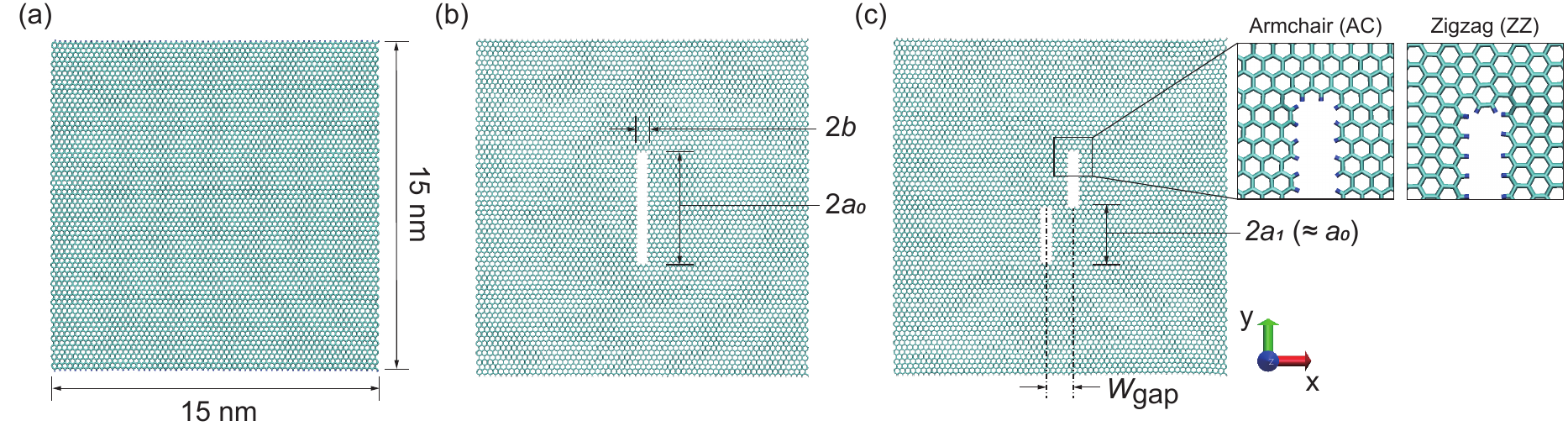}
    \caption{\label{fig:geometry-of-graphene}Geometry and structure of graphene: (a) pristine graphene, (b) graphene with a single crack of length $2a_0$ and width $2b$, and (c) graphene with two cracks, each of length $2a_1$ and width $2b$, separated by a distance $W_\text{gap}$. A magnified view of a portion of the crack is provided to show its local structure. Cyan (blue) represents carbon (hydrogen) atoms in the structure.}    
\end{figure}

Graphene is a well-known structure formed by a network of sp$^2$
carbon atoms located at the vertices of a hexagonal
lattice~\citep{Neto2009RPG}.  The atomic models of pristine graphene
and graphene with single and multiple preexisting cracks are
considered in this study, as shown in
Fig.~\ref{fig:geometry-of-graphene}. The graphene with a single crack
has a crack length of $2a_0$ and a width of $2b$, as shown in
Fig.~\ref{fig:geometry-of-graphene}(b).  The single crack is split
into two parallel cracks separated by a distance $W_\text{gap}$ to
form graphene with parallel cracks, as shown in
Fig.~\ref{fig:geometry-of-graphene}. The parallel cracks exhibit
identical crack geometry, each crack with a crack length of $2a_1$ and
a width of $2b$. The crack length $2a_1$ is not exactly half of $2a_0$
due to the atomic structure. Both armchair (AC) and zigzag (ZZ)
structures are prepared with the corresponding patterns along crack
boundaries.  The dimensions of the structures are approximately 15~nm
in width and length.  The geometry of the single crack and two
parallel cracks are listed as in Table.~\ref{tab:crack-geometry}.
Given that sample size might affect the stress–strain response, all
simulated structures are designed to be sufficiently larger than the
crack dimensions and the distances between adjacent cracks, minimizing
size effects.  The boundaries of all structures along the
$y$-direction are passivated by hydrogen, as well as the internal
boundaries of the cracks.

The tensile test is simulated on the graphene structures by applying tensile strain along the $x$-direction. This study investigates the influence of crack gap $W_\text{gap}$ on the mechanical properties and fracture behavior of the graphene structure. Pristine graphene and graphene with a single crack, each with armchair and zigzag orientations, are used as reference cases. The simulations will encompass a total of 16 structures with the crack geometry, as listed in Table.~\ref{tab:crack-geometry}.
\begin{table}[]
\captionsetup{font=normalsize}
\caption{\label{tab:crack-geometry}Graphene structures featuring preexisting cracks, as shown in Fig.~\ref{fig:geometry-of-graphene}. The labels \enquote{AC} and \enquote{ZZ} denote armchair and zigzag chiralities, respectively, and the following digit indicates the case number. Cases AC1, AC6, ZZ1, and ZZ6 have no $2a_1$ because they are structures with a single crack, representing $W_\text{gap}=0$.}
\centering
\setlength{\tabcolsep}{3.5pt}
\renewcommand{\arraystretch}{1.1}
\begin{tabular}{c c c c c| c c c c c}
\toprule
Case & $2a_0$~(nm) & $2a_1$~(nm) & $2b$~(nm) & $W_\text{gap}$~(nm) &
Case & $2a_0$~(nm) & $2a_1$~(nm) & $2b$~(nm) & $W_\text{gap}$~(nm) \\
\midrule
AC1 & 5.388 & - & 0.614 & 0
    & ZZ1 & 5.281 & - & 0.567 & 0\\ 
AC2 & - & 2.836 & 0.614 & 1.228
    & ZZ2 & - & 2.825 & 0.567 & 1.276\\
AC3 & - & 2.836 & 0.614 & 1.719
    & ZZ3 & - & 2.825 & 0.567 & 1.701\\ 
AC4 & - & 2.836 & 0.614 & 2.947
    & ZZ4 & - & 2.825 & 0.567 & 2.978\\ 
AC5 & - & 2.836 & 0.614 & 4.175
    & ZZ5 & - & 2.825 & 0.567 & 4.254\\ 
\midrule
AC6 & 2.411 & - & 0.614 & 0
    & ZZ6 & 2.333 & - & 0.567 & 0\\ 
AC7 & - & 1.347 & 0.614 & 1.719
    & ZZ7 & - & 1.351 & 0.567 & 1.701\\ 
AC8 & - & 1.347 & 0.614 & 4.175
    & ZZ8 & - & 1.351 & 0.567 & 4.254\\ 
\bottomrule
\end{tabular}
\end{table}
\subsection{\label{sec:protocol}Computational methods and protocols}

\subsubsection{\label{sec:choiceofpotential}Choice of Force Field}

Molecular dynamics (MD) simulations will be performed using the
computational tools of the LAMMPS package~\citep{lammps2022}. The
reactive force field (ReaxFF) potential~\citep{vanDuin2021} will be
used to model the interaction between carbon-carbon and
carbon-hydrogen atoms. Despite the extensive utilization of the
\enquote{adaptive intermolecular reactive empirical bond order
  (AIREBO)} potential~\citep{brenner2002,stuart2000} in the study of
the mechanical and thermal properties of carbon
nanostructures~\citep{Yakobson1996PRL,KJ2002NL,KJ2003,Rassin2010,NeekAmal2011PRB,FonsecaMuniz2015JPCC,Fonseca2021PRB},
including crack propagation and fracture of
graphene~\citep{zhao2010JAP,Zhang2012,Zhang2014,Jhon2014Carbon,Buehler2015EML,Chen2015Carbon,Yin2015NL,Shekhawat2016NatComm,Budarapu2017IJF,Lee2019Mech,FonsecaGalvao2019Carbon,felixPCCP2022,Ma2024DRM},
concerns regarding the reliability of AIREBO in accurately simulating
carbon-carbon bond breaking have been documented, even when accounting
for a specific adjustment to the \emph{cutoff} parameter of the
potential~\citep{Tangarife2019Carbon}.  As the objective of this
investigation is not to provide a comparison to the results of the
aforementioned studies, we opted to select a force field that may
prove more effective in dealing with the breaking of carbon-carbon
bonds and reactions to study the crack coalescence in graphene.

ReaxFF was developed to enable more precise simulations of chemical reactions, including bond breaking and formation~\citep{vanDuin2021}. Consequently, it is an optimal choice
of force field for the present study of the coalescence phenomenon of
cracks in graphene. 
Indeed, ReaxFF has been employed to effectively simulate the full stress-strain relationship of several one-dimensional~\citep{DeSousa2016Carbon,DeSousa2019CMS,Brandao2021CPC,Nair2011EL,DeSousa2021CP2,Brandao2023MM2} and two-dimensional (2D)~\citep{DeSousa2021CP1,DeSousa2016RSCAdv,Roman2019MM,Shishir2021MM,Brandao2023MM} nanomaterials. 

The ReaxFF set of parameters derived to simulate combustion in carbon,
hydrogen and oxygen (C-H-O) systems, developed by Chenoweth {\it et
  al.}~\citep{Chenoweth2008JPCA}, was selected for use.  This set of
parameters is included with the standard distribution of LAMMPS and
has been employed, for example, to study the mechanical and fracture
properties of graphene oxide
structures~\citep{Shenoy2010ACSNano,Bu2022ASS,Reza2023CMS} and
graphyne~\citep{RodionovJACSs2022,fonseca2025PNAS}.

\subsubsection{\label{sec:protocols1}MD methods I: boundary conditions}
The tensile simulations are performed without periodic boundary
conditions. Dirichlet boundary conditions are imposed on two narrow
regions at the leftmost and rightmost vertical edges of the structure,
each with a width of 2.4~\AA. The region on the left is kept fixed in
the \textit{x}-direction, while the region on the right is made to
move at a constant velocity. This represents the external application
of strain at a constant strain rate.  This approach eliminates the
need for barostatting algorithms, which require the correct use of
damping factors~\citep{Barret2925JCTC} and demand extensive simulation
times.

Two distinct types of Dirichlet boundary conditions are tested on
representative samples to determine the most appropriate
tensile-strain application protocol for this study.  The first type of
boundary condition involves fully constraining the atoms in the left
region and applying a constant velocity in the $x$-direction to the
atoms in the right region, while fixing their $y$- and
$z$-coordinates. This type is referred to as \enquote{Fixed BC} (where
BC denotes boundary conditions).  The second type consists of fixing
only the $x$-coordinate of the atoms in the left region, while
allowing their $y$- and $z$-coordinates to move based on interactions
between them and the atoms in the rest of the structure. The atoms in
the right region are moved at a constant $x$-coordinate velocity. This
type is referred to as “Free BC”.  Simulations of the full tensile
stress-strain of two structures (one pristine and one precracked
graphene) are performed for each type of BC to determine which type is
best for our study. In these tests, the structures are stretched along
the $x$-direction by appropriately choosing the value of the
$x$-coordinate velocity, $v_x$, to represent the strain rate value of
$10^7~\text{s}^{-1}$.

\subsubsection{\label{sec:protocols2}MD methods II: main protocols}
Three primary protocols are employed in the MD simulations. The first
two are used to address the equilibration of structures at 0~K and
300~K, respectively, while the third outlines the tensile strain
numerical experiment.

In the first protocol, energy minimization is performed using the
conjugate gradient algorithm to relax as-built atomic configurations
shown in Fig.~\ref{fig:geometry-of-graphene}. The convergence criteria
are set to 10$^{-9}$ (dimensionless) for energy and
10$^{-9}$(kcal/mol)/\AA \ for force. This initial step is essential
for correcting atomic positions according to the force field.

In the second protocol, the structures are then equilibrated at 300 K
using the following protocol.  All dynamic simulations are performed
with a timestep of 0.05~fs.  A Langevin
thermostat~\citep{Schneider1978PRB} is applied to all atoms of the
structures except those located in the leftmost and rightmost regions,
as defined in the previous subsection for the boundary conditions. The
target temperature and damping factor are set to 300 K and 5 fs,
respectively.  Atoms in the left and right boundary regions are
constrained differently. For the atoms in these regions, their
$z$-coordinates are kept fixed to resemble clamped boundaries, while
their $x$- and $y$-coordinates are allowed to freely move in
accordance with their interactions with the other atoms of the system.
This equilibration protocol ensures full relaxation of the system,
including thermal expansion effects that alleviate any residual
thermal stresses.  The equilibrium size along the $x$-direction,
$L_0$, is defined as the difference between the average
$x$-coordinates of the atoms located in the right and left boundary
regions.

The third protocol involves the numerical tensile strain experiment,
where the conditions of the type of BC (Fixed or Free) are considered.
In these simulations, the atoms on the left are kept fixed, while the
atoms on the right are moved along the $x$-direction at the constant
velocity given by%
\begin{equation}
    \label{vx}
    v_x=L_0\dot{\varepsilon} \, ,
\end{equation}
where $\dot{\varepsilon}$ is the strain rate.

LAMMPS calculates the stress tensor for every atom of the system
according to the following formula:
\begin{equation}
    \label{tensor}
    \sigma_{ab}=mv_av_b+W_{ab} \, ,
\end{equation}
where $a$ and $b$ represent any of the $x$, $y$ or $z$ coordinates,
and $W$ represents the virial contribution to the stress on one atom
due to its interaction with $N_p$ neighbors, through:
\begin{equation}
    \label{W}
    W_{ab}=\frac{1}{2}\sum^{N_p}_j\left(r_{1a}F_{1b}+r_{2b}F_{2a}\right) \, ,
\end{equation}
where $\bm{r}_j$ ($j=1,2$) are the position vectors of a pair of
interacting atoms with $\bm{F}_j$ being the resultant force on atom
$j$.

The total stress along $x$-direction is, therefore, computed from the
summation of the $x$-component of the stresses, $\sigma_{xx}$ (i.e.,
setting $a=b=x$ in eq. (\ref{tensor})), of all carbon atoms but those
at the edge regions of the system. Similar for the stresses along $y$
and $z$ directions.

In order to get a qualitative picture of the local stresses on the
structure during the tensile test, we computed the von
Mises~\citep{vonMises1913} stress per atom, $\sigma_{\mbox{vm}}$. It
is given by:
\begin{equation}
    \label{vm}
    \sigma_{\mbox{vm}}=\sqrt{0.5\left[(\sigma_{xx}-\sigma_{yy})^2+(\sigma_{yy}-\sigma_{zz})^2+(\sigma_{zz}-\sigma_{xx})^2+6(\sigma_{xy}^2+\sigma_{yz}^2+\sigma_{zx}^2) \right]},
\end{equation}
with $\sigma_{ab}$ of every atom calculated from eq. (\ref{tensor}).

The total stress and strain values of the system are exported every
500 timesteps, while the frames corresponding to the structure under
the tensile strain experiment are exported every 0.5\% strain.

A critical issue regarding the stress calculations in LAMMPS should be
discussed.  As the volume of an atom is not a well-defined quantity,
the LAMMPS algorithms that calculate the stress per atom express it in
units of [stress.volume].  To obtain values of stresses and other
mechanical quantities related to the stress-strain curve in units of
stress, the volume of the whole system must be calculated or, at
least, estimated. In the case of one-atom thick structures such as
graphene, the longitudinal, $L_x$, and transversal, $L_y$, dimensions
are relatively well defined, while the thickness, $t$, is usually
taken as equal to the distance between layers in graphite.
Consequently, to calculate the stresses of a graphene sample, it
suffices to calculate its volume by simply multiplying $L_xL_yt$.
This method for obtaining the stresses from the above LAMMPS
computation is referred to here as the \enquote{common method}.

In this study, the structures under consideration are not homogeneous
two-dimensional (2D) systems.  They contain holes, and as displayed in
the Results section, the tensile strain generates multiple undulations
and partial ruptures within the systems.  The volume of these
structures cannot be anymore accurately represented as the product of
the longitudinal, transversal, and thickness dimensions.
Consequently, rather than utilizing the external dimensions of our
structures to compute their volumes, we opted to employ another LAMMPS
algorithm that computes the Voronoi tessellation of the atomic system.
This approach provides an estimate of the volume per atom for each
system.  However, as this method is not commonly employed in the study
of stress-strain behavior in 2D materials, it is imperative to
ascertain whether this approach under- or overestimates the volume of
the system.  This will have a direct impact on the values of stresses.

The verification of the effect of the volume calculation by the
Voronoi method on the stresses of our structures is conducted as
follows.  We have simulated the stress-strain relation of a few
samples within only the linear elastic regime, where we are certain
that the volume of the structure can be simply calculated by
$L_xL_yt$.  Pristine graphene and graphene with two cracks with
$W_{\text{gap}}=0$ samples were chosen for this test.  In the case of
graphene with the cracks, due to their rectangular shape, its volume
can be simply calculated by $l_xl_yt$, where $l_x$ and $l_y$ are the
lateral sizes of the crack, and subtracted of the total volume of the
sample.  Then, we compare the values of the elastic moduli obtained
from both methods.  As demonstrated in the subsequent subsection,
these values are not in agreement. Therefore, a correction factor,
$c$, was defined to be further applied to all stress calculations with
the Voronoi method.

Summarizing, since the Voronoi method can more precisely capture the
volume of the deformed shape of tensile-strained structures, it was
used in all simulations. As this method overestimates (as shown in the
next subsection) the calculation of the volume, the parameter $c$ is
used to correct the stress values.

\subsection{\label{sec:sim-tests}Simulation tests}
In order to ascertain the most optimal conditions for executing all
simulations in the present study, three tests were conducted.  The
first test pertained to ascertaining the most suitable type of
boundary condition (BC).  The second test focuses on the effects of
the strain rate, $\dot{\varepsilon}$, and involves conducting MD
simulations of the full tensile stress-strain of a pristine structure
and a precracked structure at three distinct values of
$\dot{\varepsilon}$: $10^9, 10^8,$ and $10^7$ s$^{-1}$.  The third
test was designed to determine the parameter $c$ (defined in the
previous section) to correct the values of the stresses and elastic
moduli due to the use of the Voronoi method to calculate the volume of
the structures.  The ensuing subsections will present the outcomes of
these tests.

\subsubsection{\label{sec:boundary}Boundary condition test}
The two types of Dirichlet BC conditions mentioned in
subsection~\ref{sec:protocols1} were applied to AC graphene samples
with and without initial cracks, as shown in
Fig.~\ref{fig:geometry-of-graphene}, to investigate the boundary
effect and identify the most suitable condition for subsequent
simulations.  For this test, the crack gap $W_\text{gap}$ is set to
0.614~nm, equivalent to approximately four times the width of the
graphene regions at the edges.  The strain rate used for this test is
$10^7$ s$^{-1}$. The Voronoi method was used to determine the volume
of the system during the tensile strain simulations.

\begin{figure}
    \centering
    \includegraphics[width=0.9\linewidth]{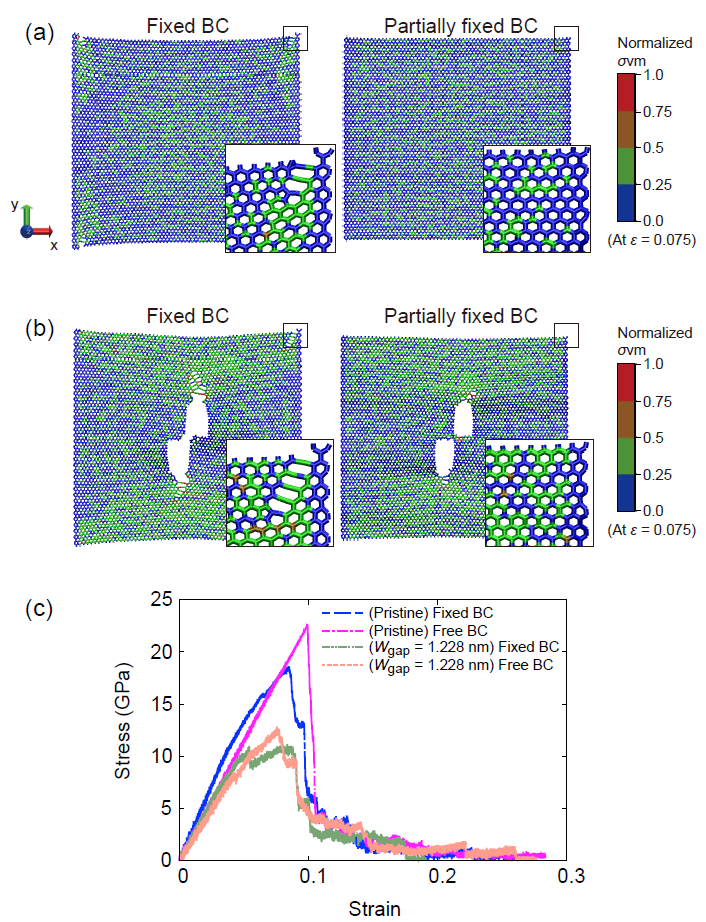}
    \caption{\label{fig:result-BCeffect}Effect of boundary conditions
      in tensile MD simulations of monolayer graphene with and without
      initial cracks at strain rate of $10^7/$s. Normalized von Mises
      stress distribution under a tensile strain of 0.06 for (a)
      pristine graphene and (b) precracked graphene. (c) Stress-strain
      curves. The correction for the stress values due to the use of
      Voronoi method was not applied here.}
\end{figure}

Fig.~\ref{fig:result-BCeffect} allows us to infer the main results of
this test. The first two rows show the pristine and precracked
graphene structures, respectively, at about 7\% of strain, for fixed
(left column) and free (right column) BCs. Different colors within the
structures represent the normalized local von Mises stresses whose
color scale is given on the right of the figures. The stress is
normalized by the certain value to maximize the visualization.  Every
structure shows a magnification of one of its corners in order to
highlight the local deformation of the atomic structure due to the
type of BC.
The fixed BC generates stresses and strains on the atoms at the
corners of the structures with high transversal components. Some
chemical bonds are broken or close to be broken. It prematurely
transmits stresses to the regions containing the cracks (compare left
and right columns of Fig.~\ref{fig:result-BCeffect}b).  Free BC does
not generate the same stresses at the edges, so the applied tensile
strain concentrates more uniformly around the cracks.  Notice that, at
the same strain, the atoms at the edges are not broken yet.

The free BC constraint enables structures to fully relax along the
transversal direction, resulting in more uniform Poisson's effect,
i.e., more uniform transversal edges during the application of tensile
strain compared to structures simulated under fixed BCs.  This
resembles the observations in tensile strain simulations performed
under transversal periodic boundary conditions, where barostat
algorithms are used to relax the system along the transversal
directions.  In contrast, the fixed BC generates a \enquote{U}-shaped
curve at the transversal edges.  This effect would likely be
negligible for larger structures.  However, larger structures would
require significantly more computational resources.  Therefore, the
free BC was employed in the simulations of the tensile strain for all
other structures.

Fig.~\ref{fig:result-BCeffect}(c) presents the stress-strain curves of
graphene structures with and without cracks, from both fixed and free
BCs.  In the case of precracked structures, although the stress-strain
curves are not significantly different when comparing fixed and free
BCs, a slight difference is observed in the inclination of the linear
region and in the maximum stress value.  However, a more pronounced
difference is observed in the stress-strain curves of pristine
graphene when comparing fixed and free BCs.  In addition to the
inclinations of the linear regions of both curves being visibly
different, the values of peak stress and tensile strain at break are
also significantly different.

Summarizing, our conclusion here is that the free BC minimizes the
influence of boundary effects on the structure of the tensile-strained
systems. In order to focus on the behavior in the cracked region, the
free BC is selected for all additional simulations.

\subsubsection{Strain rate effect}
The effect of the strain rate on the mechanical response of the system
to tensile strain is investigated for graphene with initial cracks
separated by $W_\text{gap}=1.228$~nm, as shown in
Fig.~\ref{fig:geometry-of-graphene}. The crack surface is taken in the
armchair direction. The strain rate, $\dot{\varepsilon}$, values
considered in this test are $10^7$~s$^{-1}$, $10^8$~s$^{-1}$, and
$10^9$~s$^{-1}$.

\begin{figure}[tbph]
    \centering
    \includegraphics[width=0.6\linewidth]{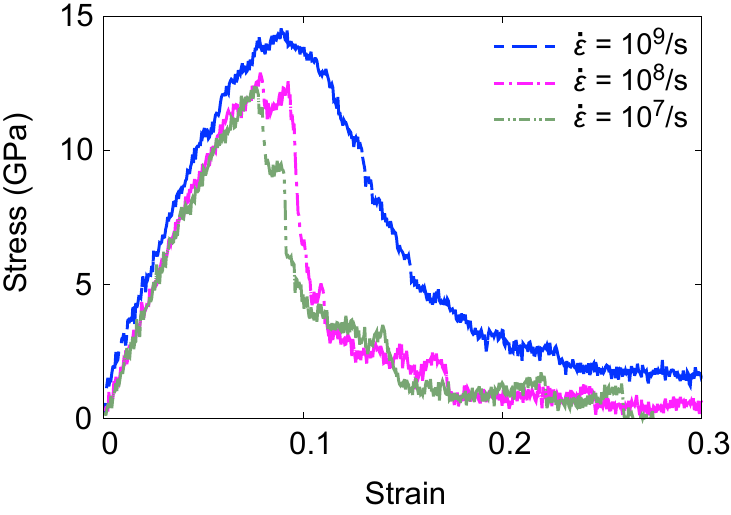}
    \caption{\label{fig:strain-effect-3AC}Stress-strain curves for
      different strain rates for the precracked graphene structure
      with $W_\text{gap}=1.228$~nm. The correction for the stress
      values due to the use of Voronoi method was not applied here.}

\end{figure}

Fig.~\ref{fig:strain-effect-3AC} shows the stress-strain curves at
varying strain rates for the system under investigation. The primary
observation is that while the peak stress increases with
$\dot{\varepsilon}$, it exhibits a substantial increase when the
strain rate is varied from $10^8$~s$^{-1}$ and $10^9$~s$^{-1}$.
Furthermore, it is evident that the responses at strain rates of
$10^7$~s$^{-1}$ and $10^8$~s$^{-1}$ are similar.  Specifically, the
peak stress values of 12.92~GPa and 13.27~GPa are observed at strain
rates of $10^7$~s$^{-1}$ and $10^8$~s$^{-1}$, respectively.  The
corresponding strain values of 0.0767 and 0.0790, recorded at these
peak stress levels, are also very close.  The difference in peak
stress is -0.35~GPa, which corresponds to a -2.71\% relative
change. Therefore, as the results for strain rates of $10^7$~s$^{-1}$
and $10^8$~s$^{-1}$ are analogous, the former is selected for all
subsequent fracture simulations.  This approach is employed to reduce
the time required for computational calculations while maintaining the
physical significance of the data.

\subsubsection{Volume correction}
As previously explained, the stress per atom, as calculated by LAMMPS,
requires the computation of the system's volume.  As the holes undergo
changes in shape and size during the tensile-strain numerical
experiments, and the plastic regime of deformation causes undulations
and partial bond breakings within the structure, the Voronoi
tessellation method is chosen because it is capable of continuously
estimating the volume of the system taking into account the various
forms of deformations observed in the different structures.  However,
it is imperative to ascertain that this method does not under- or
overestimate the volume, as this would adversely affect the stress
values.

To accomplish this, tensile simulations of pristine and AC and ZZ
graphene structures with $W_\text{gap}=0$ were conducted, using the
\enquote{common method} to estimate the volume.  This method involves
multiplying the dimensions of the structure to find its volume, which
is valid during the elastic regime.  Therefore, these tensile
simulations will be performed only for up to 2.5 \% of strain.  The
Young's moduli of these structures are extracted and compared to those
from the simulations where the Voronoi method was used to calculate
the system's volume.

\begin{figure}
    \centering
    \includegraphics[width=\linewidth]{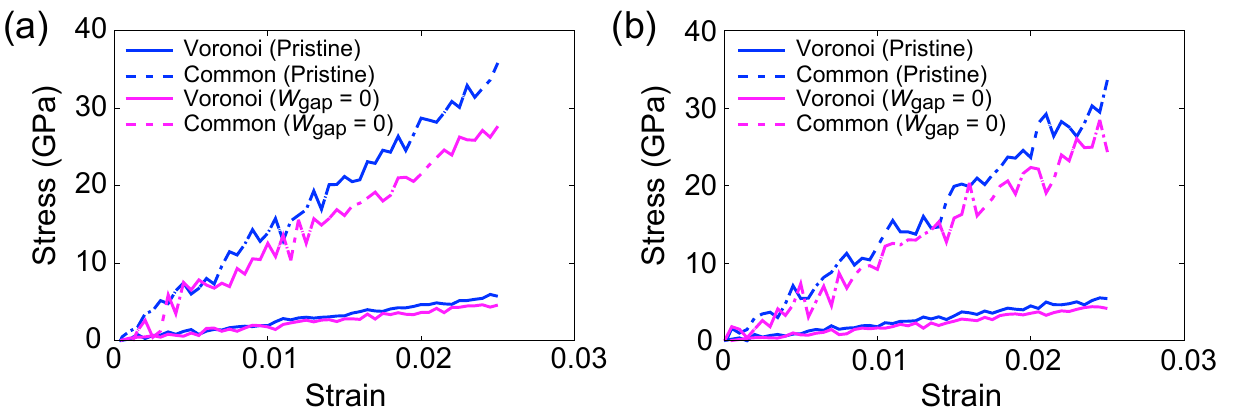}
    \caption{\label{fig:volumeMethod}Comparison of stress versus
      strain curves at the linear elastic regime, obtained by using
      Voronoi (solid line) and Common (dashed line) volume methods for
      (a) armchair and (b) zigzag structures. Pristine graphene (blue
      curves) and the graphene with the initial crack
      ($W_\text{gap}$=0, magenta curves) are examined for both
      structures.}
\end{figure}

Fig.~\ref{fig:volumeMethod} shows the stress-strain curves of the
tensile strain simulations for the structures mentioned above
calculated using the Voronoi and common volume methods.  It is evident
that the relationship between stress and strain is contingent on the
method utilized for volume calculation.  Specifically, it was
determined that the Young's moduli obtained using the Voronoi volume
method are approximately six times smaller than those obtained using
the common volume method, as indicated in
Table~\ref{tab:volumeMethod}.  Given the widely accepted value of 1
TPa for the Young’s modulus of graphene~\citep{Changgu2008}, it can be
concluded that the common volume method accurately measures the stress
values, at least at the linear regime. Therefore, we define the factor
$c$ as the ratio of Young's modulus obtained by using the common
volume method to that using the Voronoi volume method.

\begin{table}[]
\captionsetup{font=normalsize}
\caption{\label{tab:volumeMethod}Correction factor $c$ obtained from the simulations of pristine and $W_{\text{gap}}=0$ structures, at different strain rate values, $\dot{\varepsilon}$, using both Voronoi and common methods to calculate the volume of system.}
\centering
\setlength{\tabcolsep}{8pt}
\renewcommand{\arraystretch}{1.25}
\begin{tabular}{c c c c c c }
\toprule
Type & Structure & \multicolumn{1}{c}{$\dot{\varepsilon}$ (s$^{-1}$)} & Volume method & Young's modulus (TPa) & Factor $c$ \\ \midrule

Armchair & Pristine & $10^7$ & Common & 1.339 &  \multirow{2}{*}{6.032}\\ \cline{4-5}
 &  &  & Voronoi & 0.222 &  \\ 
\midrule 
 
 Armchair & $W_{\text{gap}}=0$ & $10^8$ & Common & 1.150 & \multirow{2}{*}{6.319} \\ \cline{4-5}
 & & & Voronoi & 0.182 &  \\ \midrule
 
Zigzag& Pristine & $10^7$ & Common & 1.268 & \multirow{2}{*}{6.126} \\ \cline{4-5}
 & & & Voronoi & 0.207 &  \\ \midrule
 
 Zigzag & $W_{\text{gap}}=0$ & $10^8$ & Common & 1.019 & \multirow{2}{*}{5.790} \\ \cline{4-5}
 & & & Voronoi & 0.176 &  \\ \bottomrule
\end{tabular}
\end{table}


In summary, the above tests allow us to determine the factor $c$ to
correct the stress values of all structures. So, from now on, the
stress-strain curves and all results derived from them will be shown
using the application of the average values of the factor, $c=6.067$,
obtained based on Table~\ref{tab:volumeMethod}.

\section{\label{sec:main-ss-curve} Results and discussion}

Tensile simulations are performed on graphene structures with
preexisting cracks separated by $W_\text{gap}$, as described in
section~\ref{sec:graphene-geometry}, with the protocols and conditions
detailed in sections~\ref{sec:protocol} and \ref{sec:sim-tests},
respectively.  In section~\ref{subsec:main-ss-curve}, the main results
of the simulations are presented and discussed.  In
section~\ref{sec:design-mechanical-properties}, a new model to
describe the effects of the sizes of the parallel cracks on the peak
stresses of the structures is presented and discussed.  In
section~\ref{sec:fracture-mode-prediction}, a more quantitative and
detailed discussion about the fracture behavior is presented. From now
on, the stress values shown in the figures are corrected for by the
factor $c$.

\subsection{\label{subsec:main-ss-curve}Main results}

\begin{figure}
    \centering
    \includegraphics[width=\linewidth]{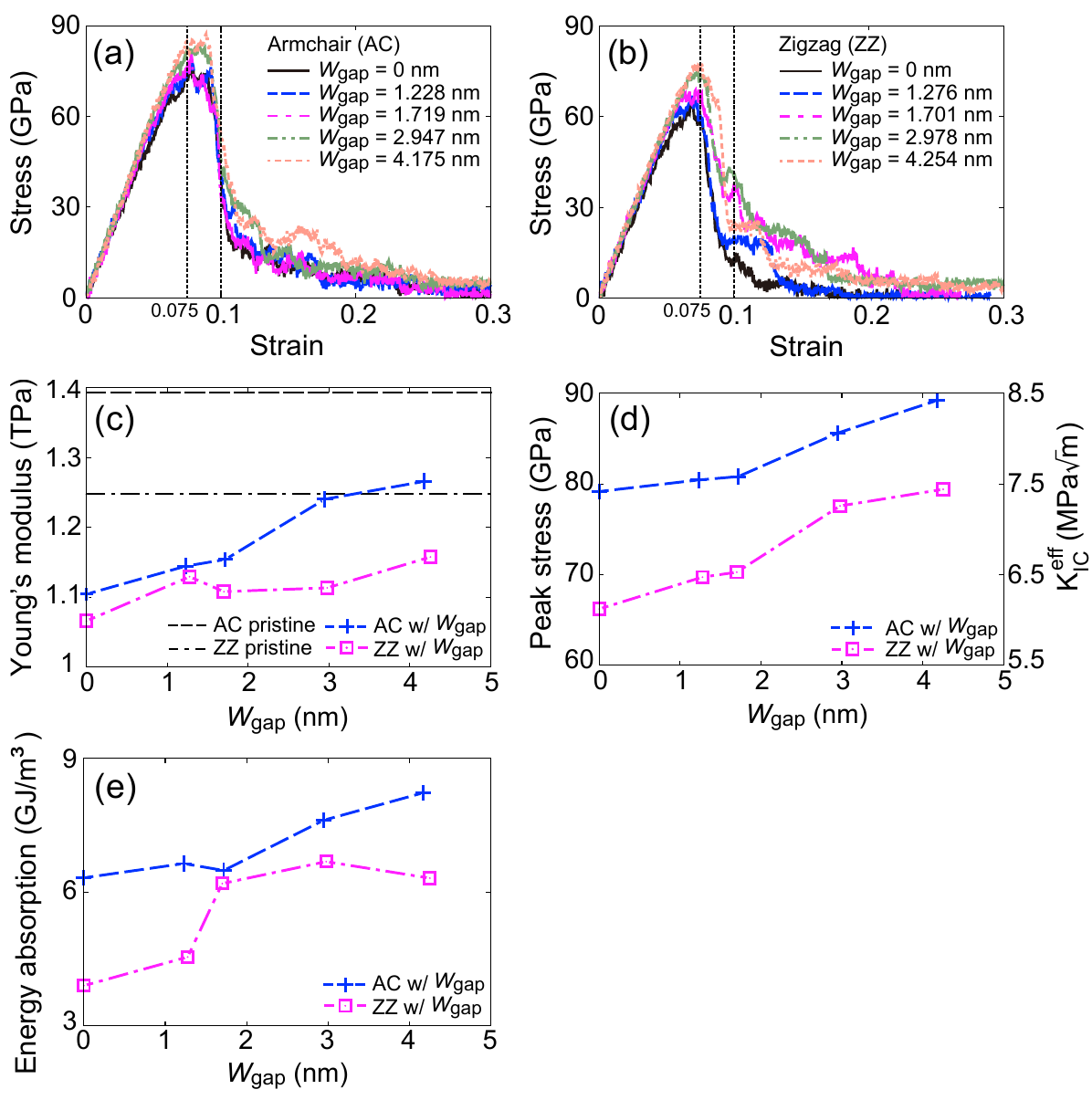}    
    \caption{\label{fig:ss-curve-main}Results of armchair (AC)
      structures with crack lengths of $2a_0 = 5.388$~nm and $2a_1 =
      2.836$~nm (AC1 to 5), and zigzag (ZZ) structures with $2a_0 =
      5.281$~nm and $2a_1 = 2.825$~nm (ZZ1 to 5). (a, b) Stress versus
      strain curves for armchair and zigzag, respectively, graphene
      structures with different values of $W_\text{gap}$, at a strain
      rate of $10^8$~s$^{-1}$. The vertical lines at strain values of
      0.075 and 0.1 are references for the next two figures. (c)
      Young's modulus versus $W_\text{gap}$ for both AC and ZZ
      structures. (d) Peak stress and effective stress intensity
      factor versus $W_\text{gap}$, respectively, corresponding to the
      stress-strain curves shown in (a) and (b). (e) Energy absorption
      under the stress-strain curve. Lines are guides to the eye. See
      the legends for the different symbols and lines.}
\end{figure}

The stress–strain curves for all values of $W_\text{gap}$ are
presented in Figs.~\ref{fig:ss-curve-main}(a) and (b) for the armchair
and zigzag structures, respectively. For these results, the armchair
structures use crack lengths of $2a_0 = 5.388$~nm and $2a_1 =
2.836$~nm (AC1 to 5), while the zigzag structures use $2a_0 =
5.281$~nm and $2a_1 = 2.825$~nm (ZZ1 to 5).  In general, the curves
for different values of $W_\text{gap}$ are very similar, although the
differences are more pronounced for ZZ structures than for AC ones.
The area under the curves also looks similar, with a perceived
tendency to increase with $W_\text{gap}$.  Although not exactly the
same, the linear region of the stress-strain curves is also quite
similar, indicating that the cracks did not significantly affect the
order of magnitude of the elastic modulus of the structures.  This is
confirmed by the data shown in Fig.~\ref{fig:ss-curve-main}(c).  The
stress-strain curves clearly show an upward shift with increasing
$W_\text{gap}$.  After the first major drop in stress, which occurs at
a strain value of about 0.1, the differences in the stress-strain
curves increase even more with $W_\text{gap}$.

More specifically, for AC structures, Fig.~\ref{fig:ss-curve-main}(a)
shows that the stress-strain curves for $W_\text{gap}=0, 1.228,$ and
1.791~nm exhibit a high degree of similarity, while the curves for
$W_\text{gap}=2.947$ and 4.175~nm demonstrate a visible upward shift.
For strain values greater than 0.1, all the curves begin to diverge
and shift upward.  Fig.~\ref{fig:ss-curve-main}(b) shows that the
stress-strain curves of ZZ structures shift upward with
$W_\text{gap}$, like the AC cases. In addition, the peak stresses in
ZZ structures have been found to be lower than those in AC structures,
which is consistent with the existing literature on the propensity for
fracture along the zigzag direction in
graphene~\citep{Jhon2014Carbon,Fujihara2015ACSNano}.  Furthermore, the
curves of ZZ structures exhibit a reduced degree of similarity between
themselves as compared to those of AC structures.  Such deviation can
be attributed to the aforementioned propensity for fracture along the
zigzag direction, in conjunction with local intercrack structure
characteristics, a subject that will be examined in the
section~\ref{sec:fracture-mode-prediction}.

The Young's modulus of the structures, calculated from the initial
response to the strain of 0.01, increases with $W_\text{gap}$, as
shown in Fig.~\ref{fig:ss-curve-main}(c). When compared with the
Young's modulus of pristine graphene, cracks reduce the Young's
modulus of graphene with defects, as expected.

The peak stress increases with $W_\text{gap}$, as shown in
Fig. \ref{fig:ss-curve-main}(d).  A quantity directly related to the
peak stress is the critical stress intensity factor,
$K^\text{eff}_\text{IC}$.  Its concept is well defined in the context
of continuum mechanics, and is related to the fracture toughness of
the materials.  Nevertheless, it can be explored in conjunction with
atomistic simulations to characterize the fracture of graphene. The
effective stress intensity factor in mode-I is defined as
\begin{equation}
    K^\text{eff}_\text{I}=\sigma\sqrt{\pi a^\text{eff},}
\label{kieff}
\end{equation}
where $\sigma$ is the stress, and $a^\text{eff}$ is the effective
crack length calculated as
\begin{equation}
    a^\text{eff}=\begin{cases}
        a_0, & \text{(single crack),} \\
        2a_1, & \text{(parallel cracks),}
    \end{cases}
\end{equation}
where the effective stress intensity factor using $a_0$ is actually
identical to the stress intensity factor.  This expression can
approximate the fracture toughness of the graphene structure with
parallel multiple cracks. The effective critical stress intensity
factor $K^\text{eff}_\text{IC}$ is, then, obtained by setting $\sigma
=\sigma^p$, where $\sigma^p$ is the peak stress of a stress-strain
curve.  Fig.~\ref{fig:ss-curve-main}(d) shows the effective critical
stress intensity factor, $K^\text{eff}_\text{IC}$, with increasing
$W_\text{gap}$.  Our values are within the range of the critical
stress intensity factors of graphene obtained from computational and
experimental data, from 3.4 to 12.0~MPa$\sqrt
{\text{m}}$~\citep{Le2016,HwaSR2014}.

The energy absorption under the stress-strain curve is also
calculated, as shown in Fig.~\ref{fig:ss-curve-main}(e). It shows that
for both AC and ZZ the energy absorption increases with
$W_\text{gap}$, demonstrating the brittle-to-ductile transition.  AC
structures are always more capable of absorbing energy than ZZ ones,
and the difference decreases for approximately $W_\text{gap}>1.3$~nm.
\begin{figure}
    \centering            
    \includegraphics[width=\linewidth]{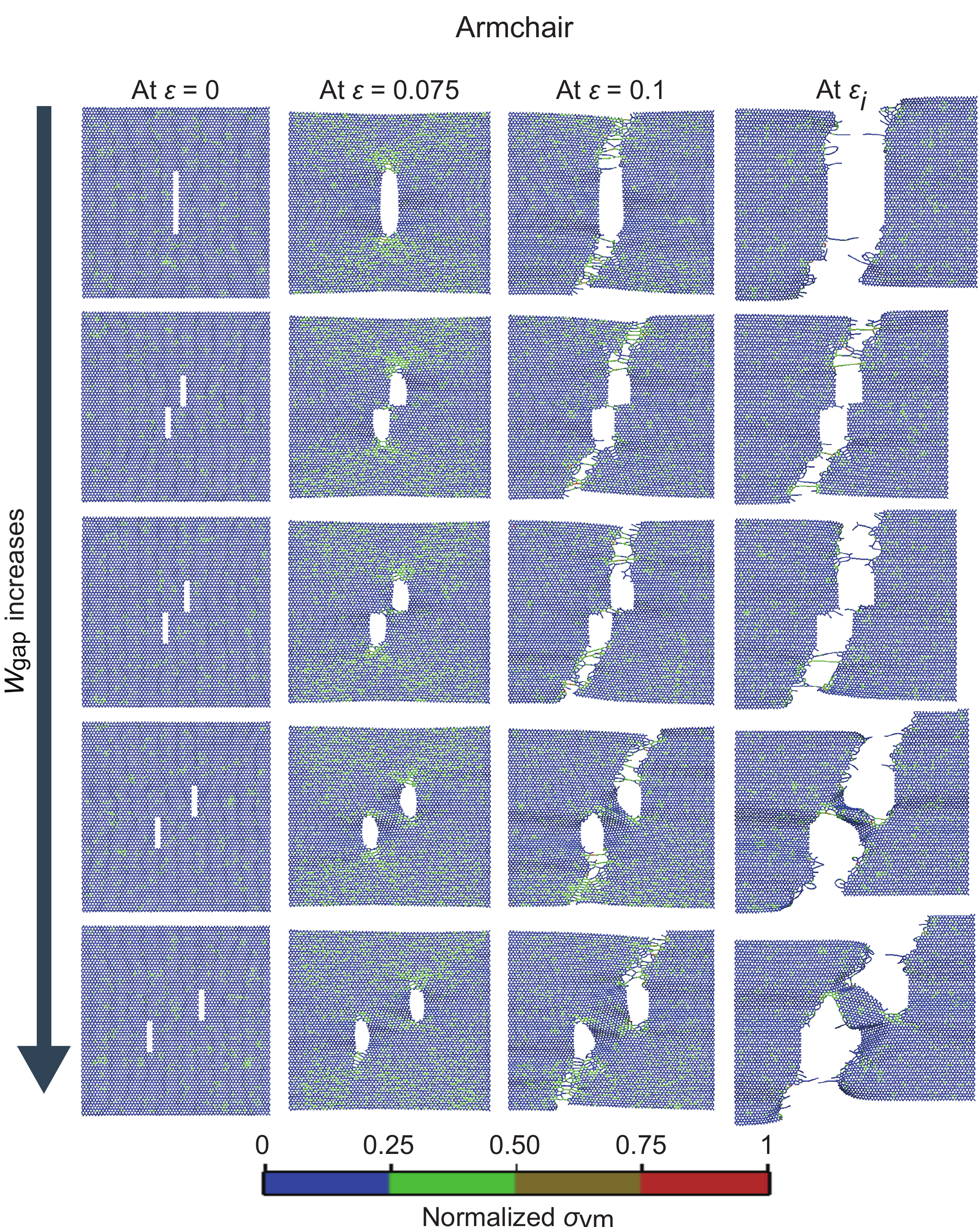}
    \caption{\label{fig:fracture-Wgap-AC} Snapshots of the AC
      structures (AC1 to 5) at certain values of tensile strain. Rows
      display structures with increasing $W_\text{gap}=0, 1.228,
      1.719, 2.947,$ and $4.175$~nm, from top to bottom. Columns
      display structures with strain values, from left to right,
      corresponding to $\varepsilon=0, 0.075,0.1$ and a specific value
      named \enquote{$\varepsilon_i$}. This last one was chosen
      according to the fracture behavior of each structure. From top
      to bottom: $\varepsilon_i = 0.28, 0.12, 0.16, 0.225$ and
      0.26. The color code corresponding to the scale of von Mises
      stresses, $\sigma\text{vm}$, is shown below the figures.}
\end{figure}
\begin{figure}
    \centering            
    \includegraphics[width=\linewidth]{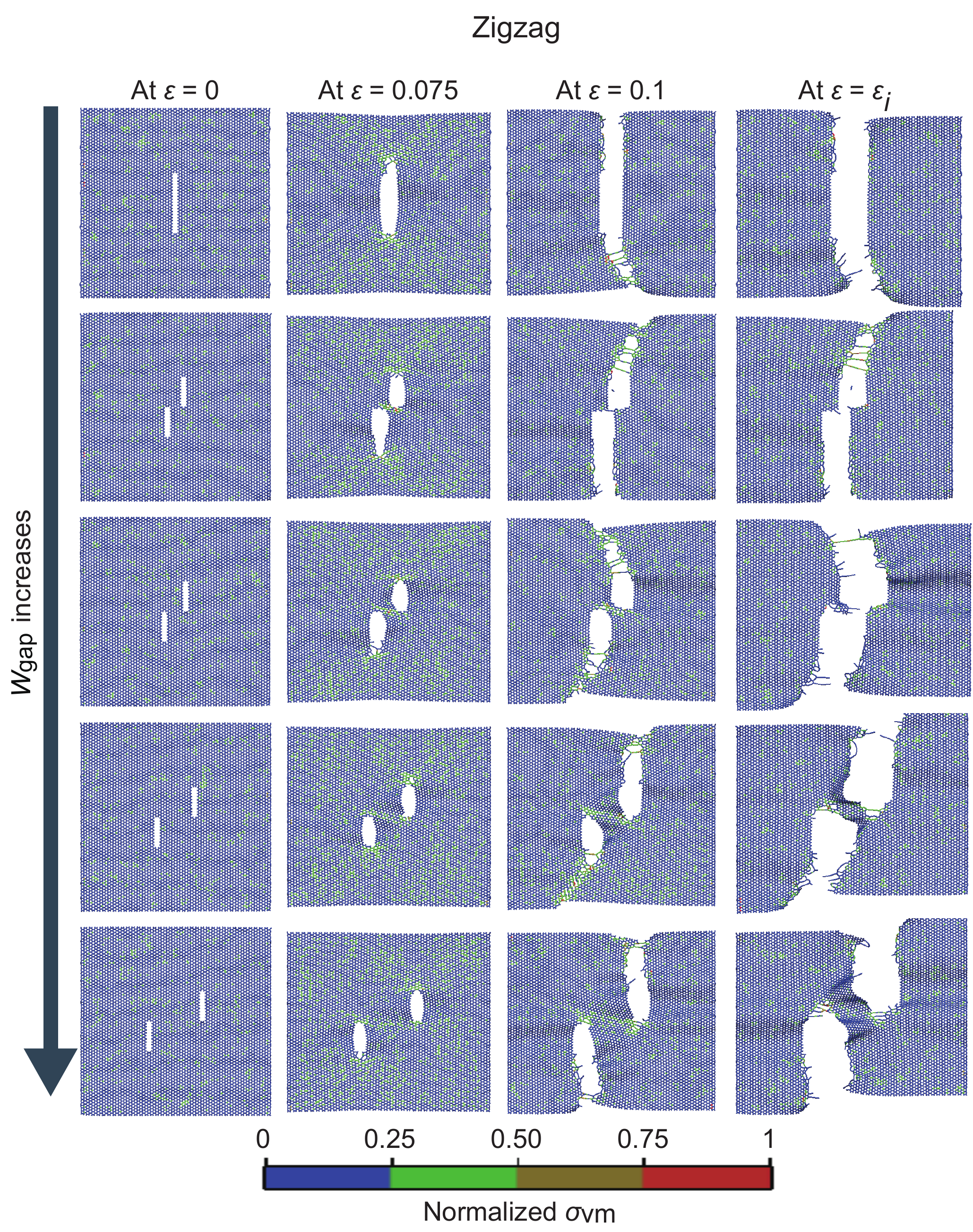}
    \caption{\label{fig:fracture-Wgap-ZZ} Snapshots of the ZZ
      structures (ZZ1 to 5) at certain values of tensile strain. Rows
      display structures with increasing $W_\text{gap}=0, 1.276,
      1.701, 2.978,$ and $4.254$~nm, from top to bottom. Columns
      display structures with strain values, from left to right,
      corresponding to $\varepsilon=0, 0.075,0.1$ and a specific value
      named \enquote{$\varepsilon_i$}. This last one was chosen
      according to the fracture behavior of each structure. From top
      to bottom: $\varepsilon_i = 0.17, 0.12, 0.215, 0.2$ and 0.2. The
      color code corresponding to the scale of von Mises stresses,
      $\sigma\text{vm}$, is shown below the figures.}
\end{figure}

Regarding the fracture behaviors of the structures, two important
general results can be observed in Figs.~\ref{fig:fracture-Wgap-AC}
and \ref{fig:fracture-Wgap-ZZ}.  They show snapshots of armchair and
zigzag structures, respectively, at strain values of 0, 0.075, 0.1 and
a particular value, named ``$\varepsilon_i$", which is chosen to
represent and illustrate the kind of fracture behavior of each
structure.  For armchair structures having $W_\text{gap}=0, 1.228,
1.791, 2.947$ and 4.175~nm, the chosen snapshots correspond to values
of $\varepsilon_i=0.28, 0.12, 0.16, 0.225$ and 0.26, respectively.
For zigzag structures having $W_\text{gap}=0, 1.276, 1.701, 2.978$ and
4.254~nm, the chosen snapshots correspond to values of
$\varepsilon_i=0.17, 0.12, 0.215, 0.2$ and 0.2, respectively.  The
strain value of $\varepsilon=0.075$ is selected for capture, as it
corresponds to the average of the strain values at peak stresses of
the structures.  The strain value $\varepsilon = 0.1$ is chosen
because, as shown in Figs.~\ref{fig:ss-curve-main}(a) and (b), at this
value, the stress either rapidly dropped or is going to drop to nearly
zero, and the structures are near fracture.  It is also observed that
the single crack propagates along the transverse direction in zigzag
structures, and along the zigzag atomic directions in armchair ones, a
feature better discussed in section
\ref{sec:fracture-mode-prediction}.

The first important result from figures~\ref{fig:fracture-Wgap-AC} and
\ref{fig:fracture-Wgap-ZZ} is that the coalescence of the cracks
occurs only for the two smallest vales of $W_\text{gap}$, for both
armchiar and zigzag structures.  The second important result regards
an interesting behavior that emerges for the structures with the two
largest vales of $W_\text{gap}$: for both armchair and zigzag
patterns, the cracks do not coalesce.  The region between the cracks,
rather than merely being strained, begins to act as a flexible lever,
causing the right side of the structure to move transversely with
respect to the left side.  This \enquote{lever} type of fracture
enables the system to withstand higher amounts of strain, which
thereby exhibits ductile behavior.  This phenomenon as well as the
details regarding the evolution of the atomic structure of the region
between the cracks during the tensile strain will be addressed in
greater detail in the subsequent quantitative discussion of failure
behaviors (section~\ref{sec:fracture-mode-prediction}).

Finally, it is worth to mention that the fracture strain also
increases with $W_\text{gap}$.  Due to thermal fluctuations, the value
of 1~GPa has been established as the stress value below which the
structure is deemed fractured for the purpose of determining the
rupture strain and calculations of toughness.

\subsection{\label{sec:design-mechanical-properties}Design of crack geometry}
Fig.~\ref{fig:ss-curve-main}(d) shows an approximately linear increase
of the peak stress $\sigma_p$ with the crack gap width,
$W_\text{gap}$.  Based on that, we propose a new, more general
relationship between $\sigma_p$ and $W_\text{gap}$, given by:
\begin{equation}
\label{eq:peak}
    \sigma_p =\sigma_p\left(a^\text{eff}, W_\text{gap}\right)= \frac{K_\text{IC}}{\sqrt{\pi a^\text{eff}}}\left(1+\alpha W_\text{gap}\right),
\end{equation}
where $K_\text{IC}$ is the critical stress intensity factor of
graphene with a single crack, and $\alpha$ is a regression coefficient
obtained from simulation data varying with $W_\text{gap}$ for a fixed
value of $a_\text{eff}$.

Eq.~\ref{eq:peak} shows the new relationship between the peak stress
and the crack gap width and effective half crack length.  To validate
this relationship, additional simulations are needed for structures
with different values of effective crack length.  They are carried out
with structures possessing $2a_1=1.347$~nm in the armchair direction
(AC6 to 8) and $2a_1=1.351$~nm in the zigzag direction (structures ZZ6
to 8).  The averaged $K_\text{IC}$ for each chirality obtained using
Eq. (\ref{kieff}) and $\sigma_p$ for each value of $a_0$, are listed
in Table~\ref{tab:KIC}.  With the results from the simulations, the
regression coefficient $\alpha$ can be obtained for each value of
$2a_1$.

\begin{figure}
    \centering
    \includegraphics[width=\linewidth]{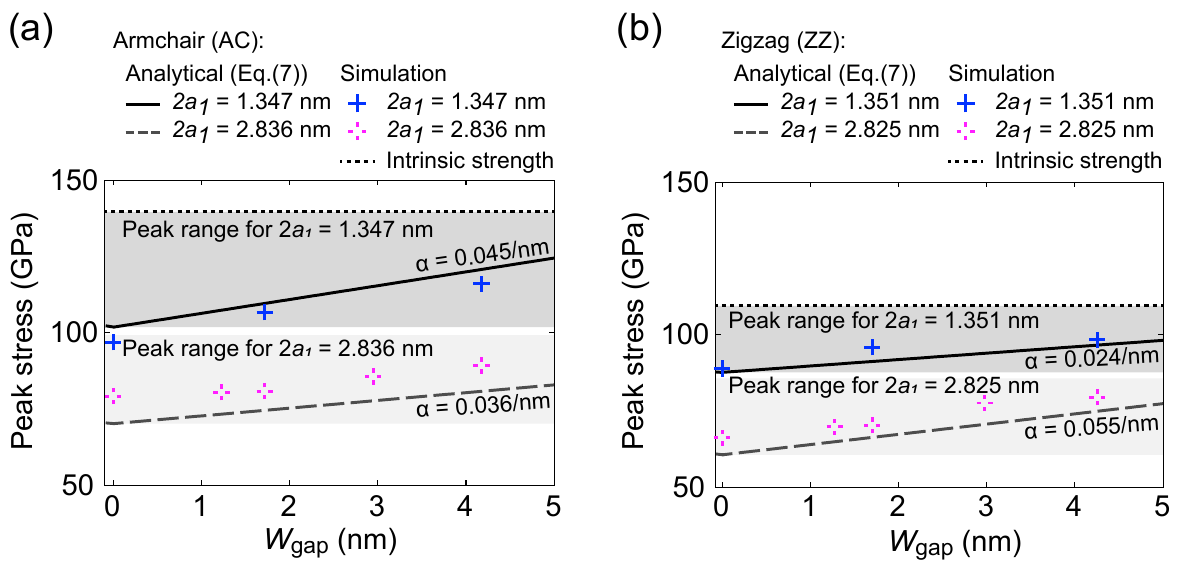}
    \caption{\label{fig:peak-design}Peak stress versus $W_\text{gap}$
      for the graphene with each crack length $2a_1$ in (a) armchair
      and (b) zigzag, respectively. The dark and light gray colored
      regions indicate the range of the possible peak stress for each
      value of $2a_1$. The value $\alpha$ is calculated for each crack
      length $2a_1$.}
\end{figure}

The excellent agreement between the analytical predictions of peak
stress based on Eq.~(\ref{eq:peak}) alongside the simulation results,
is shown in Fig.~\ref{fig:peak-design}.  It demonstrates that
$W_\text{gap}$ can be tuned to enhance the peak stress. The graph
would be symmetric with respect to the \textit{y}-axis.  The linear
model proposed in Eq. (\ref{eq:peak}) will not be valid for large
values $W_\text{gap}$, since it predicts that the peak stress can
overpass both the theoretical maximum, given by $K_\text{IC}/\sqrt{\pi
  (0.5a^\text{eff})}$, and that of pristine graphene (dotted lines in
Fig.~\ref{fig:peak-design}). Our model is interpreted as the linear
approximation of the dependence of the peak stress on $W_\text{gap}$.
Eq.~\ref{eq:peak} remains valid for crack gaps up to an upper bound
$W^\text{lim}_\text{gap}$ expressed as
\begin{equation}
    W^{\text{lim}}_\text{gap}=\frac{\sqrt{2}-1}{\alpha},
\end{equation}
derived by equating the peak stress for two cases:
$\sigma_p\left(0.5a^\text{eff},W_\text{gap}=0\right)=\sigma_p\left(a^\text{eff},W_\text{gap}\gg1\right)$. Table~\ref{tab:Wgap-limit}
shows $W^\text{lim}_\text{gap}$ corresponding to $2a_1$ for armchair
and zigzag, respectively.

\begin{table}[]
    \captionsetup{font=normalsize}
    \caption{\label{tab:KIC}Stress intensity factor $K_\text{IC}$.}
    \centering     
    \setlength{\tabcolsep}{12pt}    
    \renewcommand{\arraystretch}{1}
\begin{tabular}{c c c c}
\toprule
     Case & $2a_0$~(nm) & $K_\text{IC}$~($\text{MPa}\sqrt{\text{m}}$) & Averaged $K_\text{IC}$~($\text{MPa}\sqrt{\text{m}}$)\\
     \midrule
     AC1 & 5.388 & 7.291 & -\\     
     AC6 & 2.410 & 5.961 & -\\
     - & - & - & 6.626 \\
     \midrule
     ZZ1 & 5.281 & 6.038 & -\\
     ZZ6 & 2.333 & 5.386 & -\\
     - & - & - & 5.712 \\
     \bottomrule     
\end{tabular}
\end{table}

\begin{table}[]
    \captionsetup{font=normalsize}
    \caption{\label{tab:Wgap-limit}Design limit of crack gap $W^\text{lim}_\text{gap}$ as a function of crack length $2a_1$.}
    \centering     
    \setlength{\tabcolsep}{12pt}    
    \renewcommand{\arraystretch}{1}
\begin{tabular}{c c c c}
\toprule
     Type & $2a_1$~(nm) & $\alpha \left(\text{nm}^{-1}\right)$ & $W^\text{lim}_\text{gap}$~(nm)\\
     \midrule
     AC & 1.347 & 0.045 & 9.205 \\
     AC & 2.836 & 0.036 & 11.506 \\
     ZZ & 1.351 & 0.024 & 17.259 \\
     ZZ & 2.825 & 0.055 & 7.531 \\
     \bottomrule     
\end{tabular}
\end{table}

Another interesting result can be inferred from
Fig.~\ref{fig:peak-design}.  The coefficient $\alpha$ that determines
the linear dependence of $\sigma_p$ on $W_\text{gap}$ decreases with
increasing $2a_1$ for armchair, and increases with increasing $2a_1$
for zigzag structures. This means that the dependence of the peak
stress on $W_\text{gap}$ behaves in an opposite way between the ZZ and
AC structures.

\subsection{\label{sec:fracture-mode-prediction}Fracture analysis}
The fracture of graphene structures in this study can be classified
into three types, based on $W_\text{gap}$: single crack propagation,
crack coalescence, and separate crack propagation followed by full
rupture designated as \textit{lever} behavior.  The structures and
their corresponding fracture behaviors are listed in
Table~\ref{tab:fracture-mode-type}.  Crack coalescence is defined as
the crack connection between two already existing parallel cracks.  As
seen in Figs.~\ref{fig:fracture-Wgap-AC} and
\ref{fig:fracture-Wgap-ZZ}, when $W_\text{gap}$ increases, the
fracture mode converts from crack coalescence to lever behavior.

\begin{figure}
    \centering
    \includegraphics[width=\linewidth]{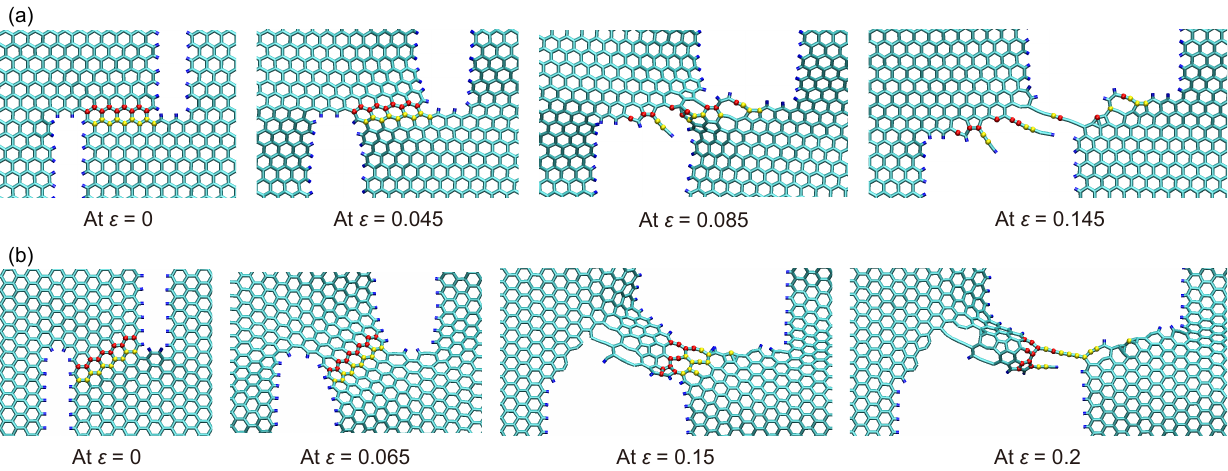}
    \caption{\label{fig:fracture-zoomin}Atomic configurations near
      crack tips for (a) armchair structure with each crack length of
      $2a_1=2.836$~nm, crack width of $2b=0.614$~nm, and gap
      $W_\text{gap}=1.719$~nm and (b) zigzag structure with
      $2a_1=2.825$~nm, $2b=0.567$~nm, and $W_\text{gap}=1.701$~nm.}
\end{figure}

Further insights into the fracture behavior are provided in
Fig.~\ref{fig:fracture-zoomin}.  It shows the atomic configurations
near the crack tips at various strain values for both armchair and
zigzag graphene structures.  Two groups of carbon atoms, colored
orange and yellow, are selected to be tracked during the tensile
strain.  The purpose is to see how the crack coalescence starts and
propagates in each type of structure/chirality.  For both chiralities,
we observe that crack coalescence initiates from the tips of the
previous cracks and progresses toward the center region as strain
increases, particularly for intermediate values of $W_\text{gap}$.  In
the case of the armchair structure with $W_\text{gap} = 1.719$~nm
(Fig.~\ref{fig:fracture-zoomin}(a)), the local atomic arrangement
exhibits a gradual shear-like distortion prior to coalescence,
indicating one kind of a progressive failure mechanism.  On the other
hand, the zigzag structure (Fig.~\ref{fig:fracture-zoomin}(b)) with a
comparable $W_\text{gap}$, exhibits more dilative deformation followed
by localized and less progressive bond breaking, which is consistent
with the lower fracture toughness observed in zigzag graphene.

Notably, in both structures, the deformation patterns suggest that the
crack tips act as localized stress concentrators where atomic bond
rotation and stretching precede rupture.  As the strain increases, the
transition from elastic deformation to crack propagation becomes
visible through increased atomic displacement and asymmetry around the
crack front.  These visualizations support the interpretation of the
\enquote{lever} behavior observed in larger $W_\text{gap}$ structures,
where the region between cracks deforms out-of-plane, allowing the
structure to sustain additional strain before failure.  Therefore,
Figure~\ref{fig:fracture-zoomin} provides atomistic evidence
supporting the shift in fracture mode from brittle (coalescence) to
ductile-like (lever) as $W_\text{gap}$ increases.

\begin{table}[]
    \captionsetup{font=normalsize}
    \caption{\label{tab:fracture-mode-type}Fracture behavior of graphene structures as $W_\text{gap}$ increases.}
    \centering     
    \setlength{\tabcolsep}{20pt}    
    \renewcommand{\arraystretch}{1}
    \begin{tabular}{c c c| c c c}
    \toprule
         Case & $W_\text{gap}$ & Fracture mode & Case & $W_\text{gap}$ & Fracture behavior\\
         \midrule
         AC1 & 0 & Single crack  
            & ZZ1 & 0 & Single crack\\
         AC2 & 1.228 & Coalescence 
            & ZZ2 & 1.276 & Coalescence\\
         AC3 & 1.719 & Coalescence
            & ZZ3 & 1.701 & Coalescence\\
         AC4 & 2.947 & Lever 
            & ZZ4 & 2.978 & Lever \\
         AC5 & 4.175 & Lever 
            & ZZ5 & 4.254 & Lever\\         
         \bottomrule
    \end{tabular}
\end{table}

\section{\label{sec:conclusion}Conclusion}
This study investigates the fracture behavior of graphene structures
with preexisting parallel cracks, focusing on the influence of crack
spacing and chirality under tensile strain.  Molecular dynamics
simulations revealed that crack coalescence occurs at smaller
$W_\text{gap}$, leading to a significant reduction in strength.
Stress-strain curves demonstrate an upward shift as the gap between
cracks $W_\text{gap}$ increases, indicating a higher peak stress and a
tendency for crack coalescence when the crack distances are small.
Crack coalescence is observed when $W_\text{gap}$ is below a specific
threshold, leading to a significant deterioration in the strength of
the material.  For larger crack separations, independent crack
propagation occurs, with the material exhibiting ductile behavior due
to a \enquote{lever} kind of distortion mechanism, highlighting the
transition from brittle to ductile fracture as the crack gap
increases.  The study also establishes that effective stress intensity
factors increase with the gap size, further indicating the correlation
between crack geometry and the mechanical properties of
graphene. Then, the design guidelines for the length and gap of
parallel cracks are discussed to predict the peak stress based on the
crack geometry.  Overall, the results suggest that the geometry and
distance between cracks play a critical role in determining the
fracture behavior of graphene. The findings offer important insights
for designing graphene-based materials with controlled fracture
properties, which could be crucial for engineering applications that
require high mechanical integrity in the presence of defects.

\section*{Acknowledgement}
This work was supported by Basic Science Research Program through the
National Research Foundation of Korea (NRF) funded by the Ministry of
Education (No. RS-2023-00242455). AFF is a fellow of the Brazilian
Agency CNPq-Brazil (\#303284/2021-8 and \#302009/2025-6) and
acknowledges grants \#2023/02651-0 and \#2024/14403-4 from S\~{a}o
Paulo Research Foundation (FAPESP). This work was supported by the
Heritage Medical Research Institute (HMRI) at Caltech and the National
Science Foundation, Center to Stream Healthcare in Place (C2SHIP),
Award No. 2052827 (C.D.). Computational resources were provided by the
High-Performance Computing Center at Caltech, the Coaraci
Supercomputer (FAPESP grant \#2019/17874-0) and the Center for
Computing in Engineering and Sciences at Unicamp (FAPESP grant
\#2013/08293-7).

\bibliographystyle{elsarticle-num}
\bibliography{main}


\end{document}